\documentclass[journal=jacsat,manuscript=article]{achemso}

\usepackage[english]{babel}
\usepackage{cmap}
\usepackage[utf8]{inputenc}
\usepackage[T1]{fontenc}

\usepackage{mathptmx}
\usepackage[scaled]{helvet}

\usepackage{textcomp}
\usepackage{setspace}
\usepackage{xcolor}
\usepackage{graphicx}
\usepackage[version=4]{mhchem}
\usepackage[section]{placeins}
\usepackage{hyperref}
\usepackage{siunitx}
\usepackage{relsize}
\usepackage{etoolbox}

\newcommand\e[1]{\ensuremath{_{\text{#1}}}}

\def\ircp{Chimie ParisTech, PSL Research University, CNRS, Institut de Recherche de Chimie Paris, 75005 Paris, France}
\def\airliquide{Air Liquide, Centre de Recherche Paris Saclay, 78354 Jouy-en-Josas, France}
\def\cambridge{Department of Materials Science and Metallurgy, University of Cambridge, 27 Charles Babbage Road, Cambridge CB3 0FS, United Kingdom}
\def\isis{ISIS Facility, Rutherford Appleton Laboratory, Harwell Campus, Didcot, Oxon OX11 0QX, United Kingdom}
\def\argonne{X-ray Science Division, Advanced Photon Source, Argonne National Laboratory, 9700 South Cass Avenue, Argonne, Illinois 60439, United States}

\author{Romain Gaillac}
\affiliation{\ircp}\alsoaffiliation{\airliquide}
\author{Pluton Pullumbi}
\affiliation{\airliquide}
\author{Kevin A. Beyer}
\affiliation{\argonne}
\author{Karena W. Chapman}
\affiliation{\argonne}
\author{David A. Keen}
\affiliation{\isis}
\author{Thomas D. Bennett}
\affiliation{\cambridge}
\email{tdb35@cam.ac.uk}
\author{Fran\c{c}ois-Xavier Coudert}
\affiliation{\ircp}
\email{fx.coudert@chimie-paristech.fr}

\title{Liquid Metal--Organic Frameworks}

\makeatletter
\patchcmd{\@maketitle}{\null}{%
  \vspace*{-3cm}
  \textcolor{blue}{%
  \begin{center}%
  \href{https://doi.org/10.1038/nmat4998}%
       {Published as: \emph{Nature Materials}, \textbf{2017}, 16, 1149--1154, DOI: 10.1038/nmat4998}%
  \end{center}%
  \vskip 1cm%
  }}{}{}
\makeatother

\begin{document}

\maketitle

\clearpage

\begin{abstract}
Metal--organic frameworks (MOFs) are a family of chemically diverse materials, with applications in a wide range of fields covering engineering, physics, chemistry, biology and medicine. Research so far has focused almost entirely on crystalline structures, yet a clear trend has emerged shifting the emphasis onto disordered states of MOFs, including ``defective by design'' crystals, as well as amorphous phases such as glasses and gels. Here we introduce a MOF liquid, a strongly associated liquid obtained by melting a zeolitic imidazolate framework (ZIF), with retention of chemical configuration, coordinative bonding modes, and porosity of the parent crystalline framework. We combine \emph{in-situ} variable temperature X-ray, \emph{ex-situ} neutron pair distribution function experiments, and first principles molecular dynamics simulations to study the melting phenomenon and the nature of the liquid obtained, focusing on structural characterization at the molecular scale, dynamics of the species, and thermodynamics of the solid--liquid transition.
\end{abstract}

\clearpage

Crystalline metal-organic frameworks have been proposed for application in a variety of situations which take advantage of their highly ordered and nanoporous structures, e.g. gas sorption and separation,\cite{mason_methane_2015, Rodenas2014, Yoon2016} catalysis\cite{mondloch_destruction_2015} and ion transport.\cite{horike_ion_2013} Inherent defects,\cite{sholl_defects_2015} structural disorder\cite{Cairns2013} and near-ubiquitous flexibility\cite{schneemann_flexible_2014} present significant challenges in the development of highly robust, selective systems from perfect crystals. However, they also present opportunities, in creating functional ``defective by design'' materials.\cite{Morris2015, bennett_interplay_2017}

Non-crystalline, or amorphous MOF systems are formed by avoidance of crystallization, or induced collapse of crystalline systems by pressure, temperature, or ball-milling.\cite{bennett_amorphous_2014} In the case of the zeolitic imidazolate framework (ZIF) family,\cite{Park2006, tian_design_2007} which adopt similar structures to inorganic zeolites and are formed from M$^{n+}$ (M$^{n+}$ = e.g. Li$^+$, B$^+$, Zn$^{2+}$) inorganic nodes connected by Im based (Im = imidazolate, C$_3$H$_3$N$_2^-$) ligands, such amorphous systems resemble the continuous random network of amorphous SiO$_2$. Recently, the formation of a molten ZIF state from a crystalline phase was observed in an inert argon atmosphere. No mass loss was observed during the formation of the liquid, which upon cooling yielded a melt-quenched glass, possessing an extended Zn--Im--Zn coordination network.\cite{bennett_hybrid_2015} Unlike reversible solid--liquid transitions in 1D or 2D coordination polymers which occur below 500\,K,\cite{umeyama_reversible_2015} melting processes in ZIFs have only been observed at high temperatures, i.e. those exceeding 700\,K.

The novelty of the liquid and glass states means microscopic insight into the mechanism of melting, and the nature of the liquid produced are of great interest when considering the generality of the mechanism in the wider MOF family. However, thus far, the narrow temperature range and poorly understood kinetics-time stability of the fleeting liquid phase have precluded information on any liquid MOF state. Particularly salient and intriguing considerations pertaining to the liquid involve (i) ``memory'' of the crystalline framework conferred by remnant framework connectivity, (ii) coordinative framework dynamics and (iii) the proximity of structure to that of an ionic or a strongly associated liquid.\cite{MacFarlane2016} The ability to form hybrid ``porous liquids'', analogous to that of the organic systems of the Cooper and James groups,\cite{giri_liquids_2015} would present a significant advance in the field, and help shift attention away from the solid state.

Motivated by the above questions and linking the MOF field to liquid, glass and polymer science, we studied the melting of ZIF-4 via experimental and computational means. The dynamic nature of the transition necessitated use of first principles molecular dynamics (FPMD) calculations, which have previously been successfully used in ionic liquid and disordered carbonate systems.\cite{Kohara2014, Corradini2016} Results were then combined with \emph{in-situ} variable temperature X-ray and \emph{ex-situ} neutron pair distribution function (PDF) experiments to yield a complete picture of the melting process, and, for the first time, an insight into the structure of a MOF-liquid.

\subsection{Structural characterization upon heating and melting}

We first studied the evolution in structure of ZIF-4, which is composed of Zn(Im)$_4$ tetrahedra linked by Zn--N coordinative bonds (Figures~1a,b), and forms a three dimensional, crystalline network containing a maximum cavity diameter of 4.9\,{\AA} (Figure~1c). It shares a topology with that of the mineral variscite, CaGa$_2$O$_4$, and a melting point of \emph{ca.} 865\,K has previously been identified,\cite{} though no atomistic modelling of experimental data has to date been performed on any MOF liquid or glass.

A sample of ZIF-4 was prepared and evacuated according to previous literature procedures,\cite{wharmby_extreme_2015} and heated to 865\,K in an argon tube furnace, before natural cooling to room temperature. Neutron and X-ray total scattering data were then collected using the GEM Diffractometer at the ISIS spallation source, and the Diamond Light Source, UK. Conversion to the respective Pair Distribution Functions (PDFs) was performed after data corrections (see Methods). Reverse Monte Carlo (RMC) modelling was subsequently performed, using as a starting model a previous Zn--Im--Zn continuous random network (CRN) configuration, arising from an amorphous, non-glass ZIF phase (i.e. one which had not passed through the liquid state).\cite{bennett_structure_2010} No changes to the network topology were necessary, and the resultant configuration is shown in Figure~1d. A good fit to the experimental structure factors was obtained (Figure~1e), which reproduces all salient features. This model represents the first obtained using experimental data on a MOF-glass.

Synchrotron X-ray diffraction data were then used to evaluate structural changes in the glass upon heating. Figure~1f shows the structure factor $F(Q)$ recorded from the glass at temperatures of 304, 778, 796 and 856\,K on heating --- while the corresponding total pair distribution functions are plotted in Figure~S1. The experimental structure factor for the glass at room temperature is largely similar to that previously reported,\cite{bennett_hybrid_2015} with no visible Bragg peaks. The PDF contains the expected peaks at \emph{ca.} 1.3\,\AA, 2\,\AA, 3\,\AA, 4\,\AA, and 6\,\AA, which correspond to C--C/C--N, Zn--N, Zn--C, Zn--N and Zn--Zn pair correlations respectively.

Upon heating from 304 to 778\,K, both intensity and position of the first sharp diffraction peak (FSDP), centered at 1.1\,\AA$^{-1}$ remained approximately constant, as was the case with other visible features in the $F(Q)$. However, further heating to 856\,K resulted in a more pronounced intensity reduction and shift in the position of the FSDP to 1.3\,\AA$^{-1}$, along with a near-total disappearance of any features at higher $Q$ values. This is in stark contrast to the case of liquid silica, where negligible changes in the FSDP upon melting are indicative of significant intermediate range order.\cite{mei_structure_2007} The changes in the $F(Q)$ result in a significantly decreased Zn--Zn correlation peak in the corresponding high temperature PDFs, centered on 6\,{\AA} (Figure~S1).

To probe the evolution of the ZIF structure upon heating and liquid formation from a microsopic point of view, we performed first principles molecular dynamics simulation (FPMD) by running constant-temperature MD simulations at temperatures of 300, 600, 800, and up to 2,250\,K.\cite{note_temperatures} Because of the computational cost of FPMD, these simulations cannot be performed directly on the glass model --- the unit cell of which is prohibitively large. Instead, we used the ZIF-4 crystalline phase as starting point, with the change in structure factor upon heating in good agreement with the trends observed experimentally (Figure~1f). Similarly close agreement is also witnessed in the variable temperature total pair distribution functions (see Figure~S2).

Moreover, in addition to the total PDFs, we calculated from the MD trajectories the PDFs for specific atom-atom correlations, which provide greater understanding of the salient real-space structural movements. We plot in Figure~2a--c the partial radial distribution functions $g_{ij}(r)$ for Zn--N, Zn--Im (where Im is the center of mass of the imidazolate group), and Zn--Zn pairs. In addition to the overall thermal broadening of the peaks, there is a clear loss of long-range order at high temperature. At intermediate temperatures (around 1,000\,K), the system has liquid-like disorder, with $g_{ij}(r)$ that do not go to zero after the first peak, though it retains some order at distances larger than 10\,{\AA}. To characterize further this state, we plot in Figure~2d the generalized Lindemann ratio,\cite{chakravarty_lindemann_2007} computed from the width of the first peak in the Zn--N and Zn--Zn partial radial distribution functions. The usual criterion used to determine melting from the Lindemann ratio is between 10 and 15\%, which indicates in our case a melting temperature between 1,000 and 1,500\,K --- and we note that the Zn--Zn ratio shows a clear disruption in slope at 1,200\,K.

\subsection{Thermodynamics of melting}

Approaching the phenomenon from a thermodynamic standpoint, we plot in Figure~2e the evolution of heat capacity $C_V$ as a function of temperature. We see a clear jump in the heat capacity, indicative of a solid--liquid phase transition, from the value of the crystal phase ($C_V \approx 2.2$~J.g$^{-1}$.K$^{-1}$) to a higher value for the ZIF liquid ($C_V \approx 2.8$~J.g$^{-1}$.K$^{-1}$). Integration yields an estimate for the enthalpy of fusion of the ZIF of $\Delta H\e{fus} = 173$~J.g$^{-1}$, which is in line with values for materials such as quartz ($\Delta H\e{fus} = 146$~J.g$^{-1}$) and cristobalite ($\Delta H\e{fus} = 237$~J.g$^{-1}$).\cite{kelly1936heats}

Since the melting of this supramolecular network is dependent upon partial dissociation and reassociation of Zn--N coordination bonds, we used this distance as a reaction coordinate. From the Zn--N partial radial distribution function, $g\e{Zn--N}(r)$, we calculated the potential of mean force (PMF) between the two atoms at all temperatures, through the relation $F(r) = -k\e{B}T \ln g(r)$. From the resultant free energy profiles (Figure~S3), we were then able to extract the temperature dependence of the activation free energy ($\Delta F^\ddagger$) needed to break the Zn--N bond. It can be seen on Figure~S4 that $\Delta F^\ddagger$ follows a van 't Hoff law, $\Delta F^\ddagger(T) = \Delta H^\ddagger - T\Delta S^\ddagger$, with $\Delta H^\ddagger \approx 121$~kJ.mol$^{-1}$ and $\Delta S^\ddagger \approx 34$~J.mol$^{-1}$.K$^{-1}$ constant in this temperature range.

The main contribution to the free energy barrier is thus energetic in nature, accompanied by a minor entropic stabilisation. Moreover, the free energy barrier is still relatively high at the melting temperature, with $\Delta F^\ddagger(T=1,000\text{\,K}) = 87\text{~kJ.mol$^{-1}$} = 10.5\,k\e{B}T$. Hence, as in conventional solids, melting occurs as a rare barrier-crossing event.\cite{Samanta2014} Finally, comparison of the potential of mean force for the Zn--Im coordinate with the PMF for Zn--N shows that breaking of the Zn--N bond is indeed an entirely suitable choice of reaction coordination for this activated process: as shown in Table~S1, the thermodynamic parameters are almost identical in both cases.

\subsection{Microscopic mechanism}

Turning to a molecular level visualization of the melting process, we depict in Figures~3a and 3b the distribution of Zn cation coordination numbers, as a function of temperature. The ideal four-fold coordination is maintained at low temperatures up to 1,200\,K, where more than 94\% of the Zn ions are coordinated by four imidazolate groups. In this regime, the under-coordination of Zn$^{2+}$ can be seen as a defect in the solid, and its concentration is found (as expected, Figure~S5) to be proportional to $n\e{d} \propto \exp(-\varepsilon/k\e{B}T)$, where $\varepsilon \approx 56$~kJ.mol$^{-1}$ is the energy required for defect formation. These ``undercoordinated'' zinc ions can act as nucleation sites for melting. Going to higher temperature, the proportion of undercoordinated Zn$^{2+}$ increases dramatically, e.g. at 1,500\,K with 59\% 4-fold coordinated, 39\% in 3-fold coordination, and 2\% in 2-fold coordination. In contrast, we note that pentacoordination is almost nonexistent in our simulations.

Focusing on typical individual linker exchange events of zinc cations provided the mechanistic picture shown in Figure~3d (snapshots from a FPMD simulation at 1,500\,K): the melting process takes place over a few picoseconds through a sequence of well-defined steps. From an initial four-fold coordinated zinc, one imidazolate linker moves away and is then replaced by a neighboring Im group with a dangling nitrogen lone pair. Inspired by Laage \emph{et al.}'s statistical treatment of the molecular mechanism of reorientation in liquid water,\cite{laage_molecular_2006, laage_molecular_2008} we averaged over all events during a molecular dynamics trajectory to plot the average distance of the outgoing and incoming N atoms during an exchange --- taking the reference time $t=0$ when both nitrogren atoms are at equal distance from the zinc ion. This plot (Figure~3c) shows that the event is concerted and that the exchange itself is rather fast, lasting less than 2\,ps. This is very similar to what is seen for a switch between hydrogen bond partners in liquid water, showcasing the similarities between this and the ZIF liquid: both are strongly associated liquids forming a dynamic network with preferred tetrahedral association.

We then calculated the frequencies at which the Zn--N bonds or Zn--Im linkages break. Plotting the log of these frequencies against inverse temperature (see Figures~S6 and S7) demonstrates the Arrhenian behaviour of the system, reinforcing the idea that melting is driven by rare events disturbing the network. We also note that the two activation energies for Zn--N and Zn--Im are very close. Extrapolating to the melting temperature --- where in the limited time window of our simulations we cannot directly observe enough of these rare events to gather good statistics --- shows that the timescale at which bond breaking occurs becomes sub-nanosecond near the melting temperature. As Table~S2 shows, for any given zinc cation, we expect one coordinative bond breaking every 143\,ms at 300\,K, every 58\,ns at 600\,K, and every 0.87\,ns at 840\,K. Moreover, looking at these events in detail, we see that at low temperature, the majority of bond-breaking events leads to a simple flip of the imidazolate linker, and do not result in exchange between two linkers (Table~S3). As temperature increases, this ``local'' motion of lower energy becomes dominated by events which lead to linker exchange, and thus allows the ZIF to melt. Finally, we see that the duration of the coordinative bond breaking and reformation is independent of temperature, and is only guided by the timescale for approach of another imidazolate partner and local dynamics.

\subsection{Characterization of the liquid ZIF}

To characterize the dynamics of the ZIF in the liquid phase, we calculated from our FPMD the translational diffusion of both zinc cations and imidazolate anions. The plots of mean square displacement over time, shown in Figure~4a and 4b, clearly show diffusive behaviour at temperatures above 1,200\,K, i.e. in the liquid phase. The translational diffusion of zinc and imidazolate are clearly linked strongly, as the coefficients are very similar (see Table~S4): for example, at 1,500\,K we have $D\e{Im} = 7.7\ 10^{-10}$~m$^2$.s$^{-1}$ and $D\e{Zn} = 6.5\ 10^{-10}$~m$^2$.s$^{-1}$. This is yet another sign of the strongly associated nature of the liquid. Moreover, the translational diffusion coefficients over temperature follow an Arrhenius law (see Figures~S8 and S9), compatible with a jump-like  diffusion of ``partner exchange'' events.\cite{laage_molecular_2008} The activation energies for Zn and Im respectively are found to be 105\,kJ.mol$^{-1}$ and 102\,kJ.mol$^{-1}$ respectively. The rotational diffusion of imidazolate cycles can also be measured, where we choose to follow rotation around the N--N axis and the associated angle $\theta$. Orientational diffusion happens at lower temperatures than translational diffusion, starting around 800\,K, and can be associated with an activation energy three times lower ($\approx 40$~kJ.mol$^{-1}$). At intermediate temperatures, before melting, there thus exists a regime of free rotation of the imidazolate linkers.

Given the focus on porosity in MOFs, we investigated the nature of the liquid ZIF from a structural point of view. We performed statistical analysis of the instantaneous porosity along FPMD trajectories at all temperatures --- using a geometric criterion for the determination of porosity and a probe diameter of 2.4\,\AA, corresponding to the kinetic diameter of helium. The evolution of the pore volume distribution is depicted in Figure~4c, from 300\,K to higher temperatures. As expected for the solid phase ($T < 1,200$~K), a slight broadening of the distribution is observed, which corresponds to increased thermal motion. However, at higher temperatures and particularly in the liquid phase, we see that the porosity is maintained overall, with only a slight deviation in average to lower pore volumes. This result, obtained in constant-volume simulations performed at densities inferred from the experimentally available data, were confirmed by shorter constant-pressure simulations. We thus conclude that, even at very high temperatures, the ZIF forms a hybrid ``porous liquid'', quite different in nature from the organic systems recently reported,\cite{giri_liquids_2015} which are formed from cage molecules providing a well-defined pore space in solvents whose molecules are too bulky to enter the pores.\cite{OReilly2007, Hasell2016} This finding is in agreement with the available experimental data on the free volume of ZIF glasses.\cite{thornton_porosity_2016}

To link the predicted liquid structure to experimental information, we performed Reverse Monte Carlo (RMC) modelling on the X-ray total scattering data collected at 856\,K (Figure~1f). The atomistic configuration derived for the glass in Figure~1d was used as a starting model, with a reduced density to reflect the changes upon melting. The final configuration is shown in Figure~4d, along with the fit to the experimental X-ray structure factor (Figure~S10). Whereas the internal surface of the glass at ambient temperature was calculated to be 4.8\%, again using standard probe diameter of 2.4\,\AA, that of the liquid at 856\,K increased to 16.2\%. Whilst transient in nature, linked voids appear irregularly distributed throughout the configuration.

\subsection{Perspectives}

In this work, we introduce the general term ``MOF liquid'', for a liquid formed from the melting of a MOF, due to the retention of chemical configuration and coordinative bonding modes between the solid and liquid phases. Importantly, we show the retention of porosity in the liquid state, with a pore volume larger than in the glass state, making liquid ZIF-4 a rare example of an intrinsically porous liquid. The demonstration and rationalization of melting in ZIF-4 provides a prototypical example in this new area, and a foundation on which further studies can be based, thus opening the way to the identification (or design) of MOFs with reduced melting temperatures and increased porosity. Chemically tunable MOF liquids with tailored properties reminiscent of crystalline structures will prove accessible through modification of inorganic centers and linker functionalization, length and shape --- similar to those routinely used in the design and synthesis of MOF crystal structures. Such materials would be of interest for liquid phase separations, homogeneous catalysis and ion transport, alongside as an intermediary state in the formation of mechanically and thermally stable MOF glasses retaining porous characteristics.

%%%%%%%%%%%%%%%%%%%%%%%%%%%%%%%%%%%%%%%%%%%%%%%%%%%%%%%%%%%%%%%%%%%%%%%%%%%%%%%%%%%%

\clearpage

\begin{center}
\includegraphics[width=\linewidth]{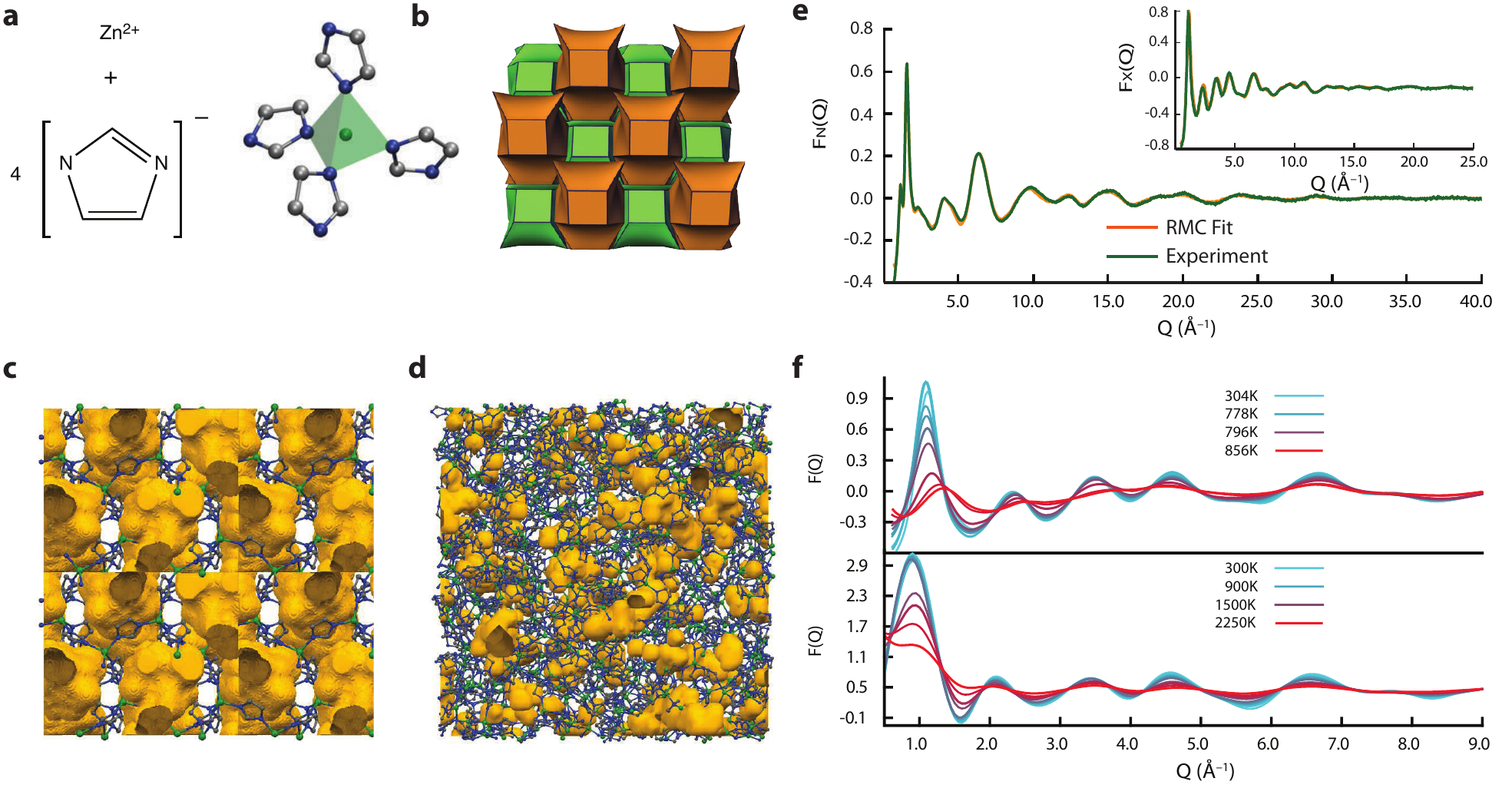}
\end{center}
\bigskip
\textbf{Figure 1. Structure of the ZIF-4 crystal, glass, and structural evolution upon heating.} (\textbf{a}) The construction from metal ion and linker of Zn(Im)$_4$ tetrahedra, the basic building unit of ZIF-4 (Im = imidazolate; Zn: green, N: blue, C: grey). (\textbf{b}) Representation of the cag topology adopted by ZIF-4, where each polyhedra corner corresponds to one Zn(Im)$_4$ tetrahedron. (\textbf{c}) Crystalline structure of ZIF-4, with free volume represented in orange. (\textbf{d}) Atomic configuration of the melt quenched glass, gained from modelling synchrotron and neutron total scattering data. (\textbf{e}) Experimental neutron structure factor $F(Q)$ data and the fit from the configuration shown in (\textbf{d}). Inset: X-ray data and fit. (\textbf{f}) Experimental glass (top) and computational ZIF-4 (bottom) X-ray structure factors upon heating.

\clearpage

\begin{center}
\includegraphics[width=\linewidth]{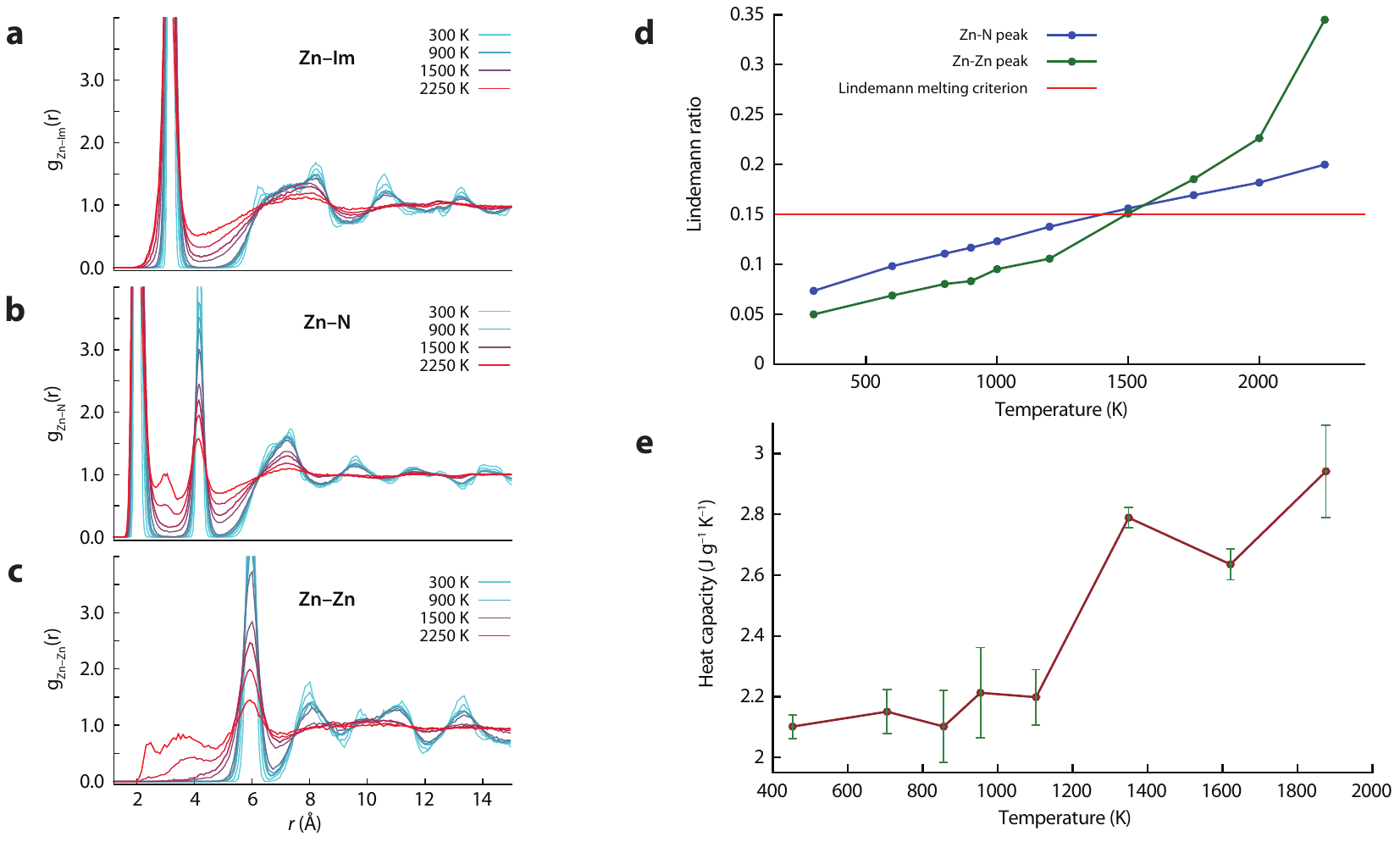}
\end{center}
\bigskip
\textbf{Figure 2. Computational data from ZIF-4 melting: structure and thermodynamics.} (\textbf{a}--\textbf{c}) Evolution of the partial radial distribution function $g_{ij}(r)$ for (\textbf{a}) Zn--Im distances (where Im is the center of mass of the imidazolate group), (\textbf{b}) Zn--N distances, (\textbf{c}) Zn--Zn distances, at temperatures going from 300\,K (light blue) to 2,250\,K (red). (\textbf{d}) Generalized Lindemann ratio $\Delta$, quantifying the liquid nature of the system, as a function of temperature, calculated for Zn--Zn (green) and Zn--N (blue) interatomic distances. The red horizontal line represents the ``critical ratio'' indicated in the literature at 10\% or 15\% (the value chosen here). (\textbf{e}) Evolution of the heat capacity with temperature.

\clearpage

\begin{center}
\includegraphics[width=\linewidth]{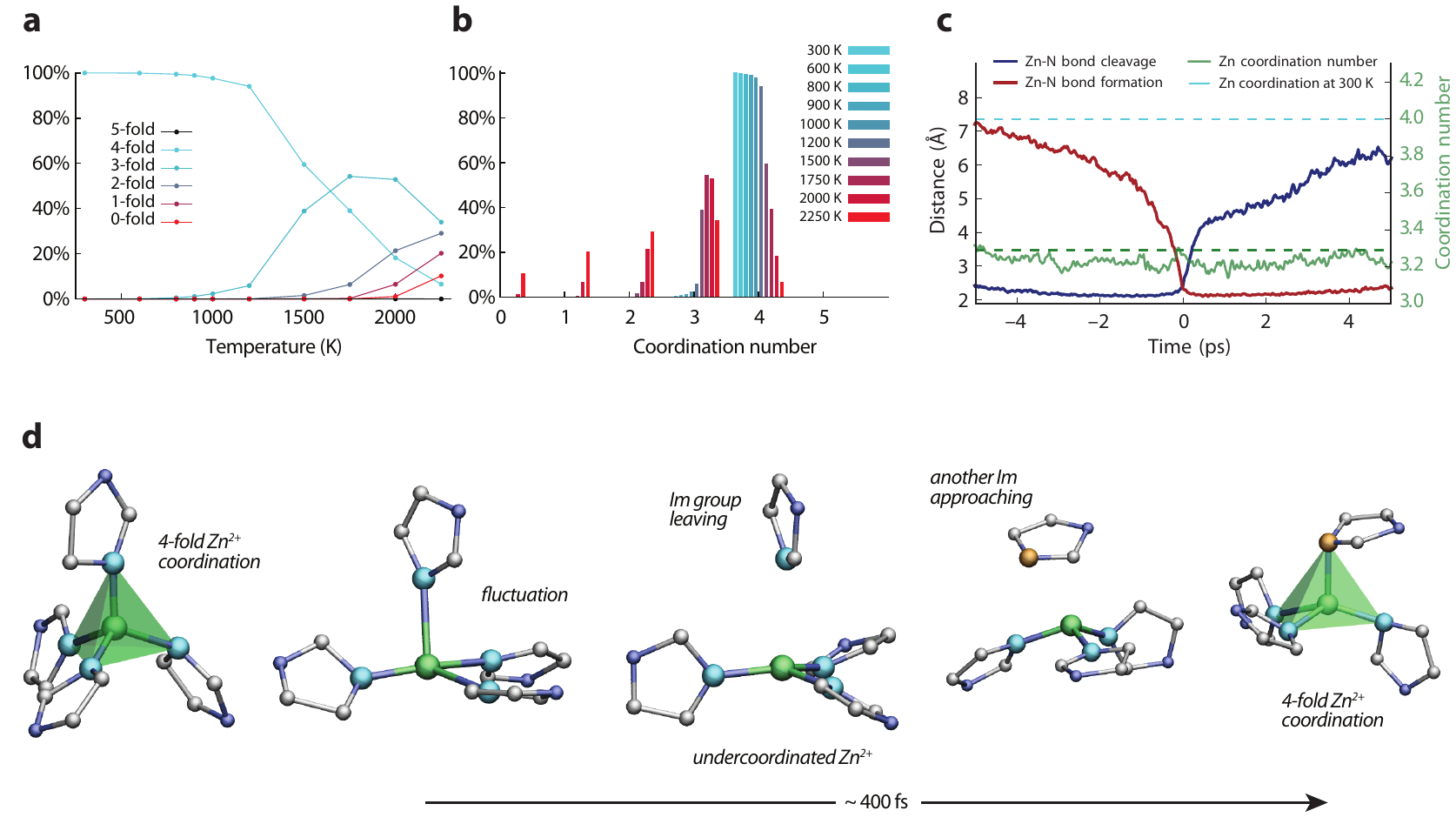}
\end{center}
\bigskip
\textbf{Figure 3. Molecular mechanism of ZIF-4 melting.} (\textbf{a}) Distribution of zinc coordination numbers as a function of temperature, from 0-fold coordinated (red) to 4-fold (light blue). 5-fold coordination is indicated in black, but is close to the temperature axis. (\textbf{b}) Temperature evolution for each degree of coordination of zinc cations. Temperatures range from 300\,K (light blue) to 2,250\,K (red). (\textbf{c}) Behaviour during an exchange of a nitrogen atom by another nitrogen atom, in the first coordination sphere of a zinc cation, averaged over all such events (all exchanges on all zinc cations). The distance between the incoming nitrogen and the zinc is plotted in red, and that between the outgoing nitrogen and the zinc in blue. The green curve corresponds to the average coordination number of the zinc cation involved in the exchange. The flat dashed lines are the average coordination number over the whole simulation at 300\,K (light blue) and 2,000\,K (green). (\textbf{d}) Visualization of a representative imidazolate exchange event. Zn: green, N (initially coordinated): light blue, N: blue, N (coordinated after exchange): orange, C: grey.

\clearpage

\begin{center}
\includegraphics[width=\linewidth]{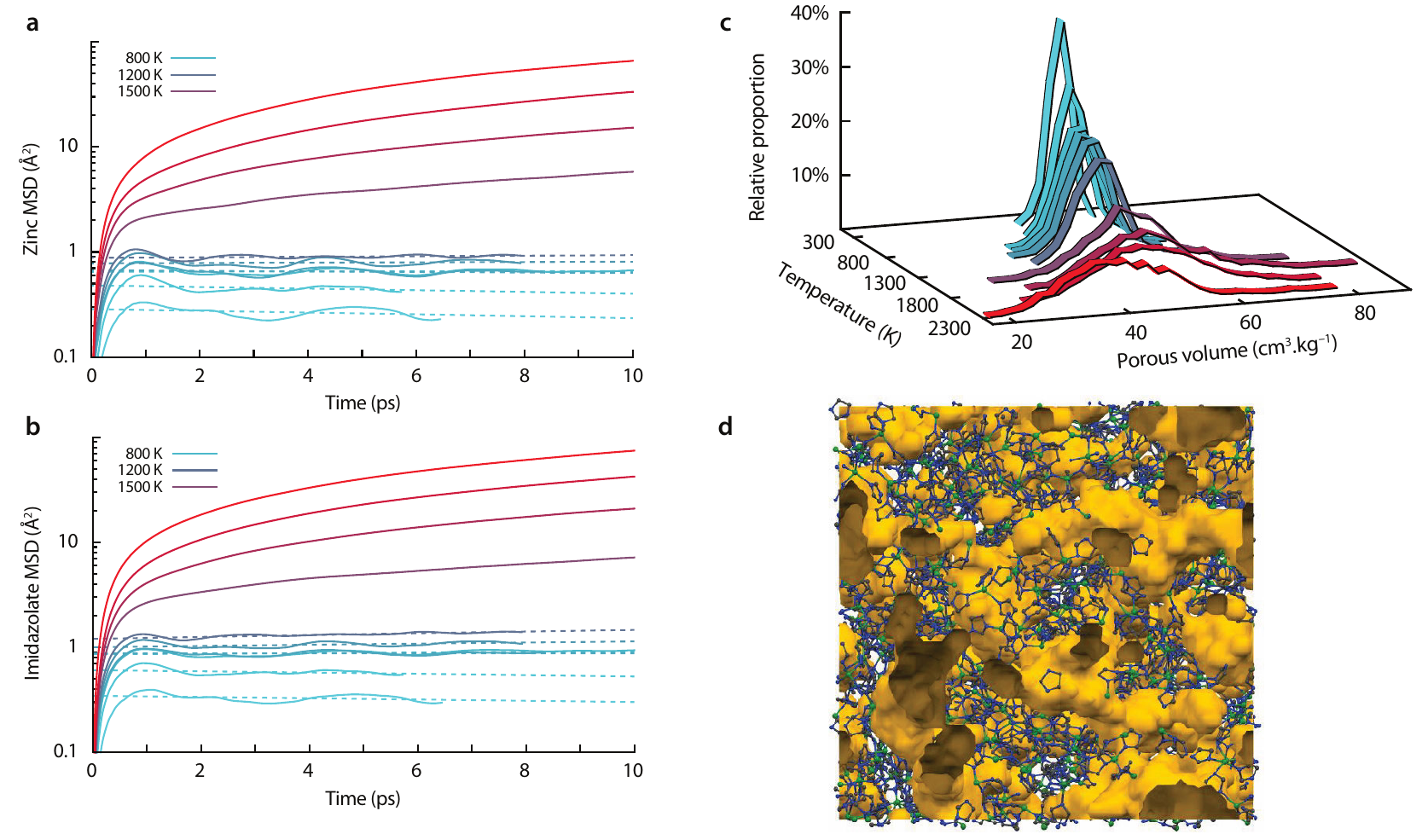}
\end{center}
\bigskip
\textbf{Figure 4. Structure and dynamics in the ZIF liquid.} (\textbf{a}-\textbf{b}) Mean square displacement (MSD) as a function of time for (\textbf{a}) zinc cations and (\textbf{b}) the center of mass of the imidazolate linkers, plotted in logarithmic scale, for temperatures ranging from 300\,K (light blue) to 2,250\,K (red). Dashed curves represent fits of the MSD (excluding short times) for lower temperatures, showing the absence of diffusive behaviour. (\textbf{c}) Temperature evolution of the distribution of the total pore volume, determined for a standard probe of radius 1.2\,{\AA}. The average pore volume takes the following values: 52\,cm$^3$.kg$^{-1}$ at 300\,K, 49\,cm$^3$.kg$^{-1}$ at 2,000\,K, and 41\,cm$^3$.kg$^{-1}$ at 2,250\,K. (\textbf{d}) Atomic configuration of the ZIF melt, gained from Reverse Monte Carlo modelling of the total scattering data collected at 856\,K. Free volume is represented in orange, Zn atoms in green, N in blue, and C in grey.

\clearpage

\section{Methods}

\subsection{Neutron Total Scattering}

A sample of deuterated ZIF-4 was prepared by directly substituting C$_3$D$_4$N$_2$ for C$_3$H$_4$N$_2$ in the synthetic methodology previously reported for ZIF-4,\cite{wharmby_extreme_2015} and melt-quenched according to published procedures.\cite{bennett_hybrid_2015} Data were measured at room temperature using the GEM diffractometer at ISIS\cite{keen_comparison_2001} on the sample which filled a 6\,mm diameter thin-walled cylindrical vanadium can to 36\,mm. Suitable background subtraction and corrections were performed using measurements from an empty vanadium can, empty instrument, 8\,mm V-5.14\% Nb rod, and the Gudrun software.\cite{keen_comparison_2001}

\subsection{X-ray Total Scattering (Room Temperature)}

Data were collected at the I15-1 beamline at the Diamond Light Source, UK ($\lambda = 0.158345$\,\AA, 78.3\,keV). A small amount of the sample used in the neutron total scattering experiment was loaded into a fused silica capilliary of 0.76\,mm diameter. Data on the sample, empty instrument and capillary were collected in the region of $\sim 0.4 < Q < \sim 26$~\AA$^{-1}$. Background, multiple scattering, container scattering, Compton scattering and absorption corrections were performed using the GudrunX program.\cite{Soper2011, soper_extracting_2011}

\subsection{X-ray Total Scattering (Variable Temperature)}

Measurements were performed on a sample of melt-quenched ZIF-4 glass produced in an identical manner to those used for the X-ray room temperature measurements. Data were collected at the Advanced Photon Source, USA on the 11-ID-B beamline ($\lambda = 0.143$~\AA, 86.7\,keV), in the range $0.6 < Q < \sim 24$~\AA$^{-1}$. A finely ground sample of the glass was loaded into a 1\,mm diameter silica capillary, along with glass wool to hold it in place during the melting process. Data were collected under flowing argon gas at room temperature, and then upon heating from 298\,K in ca. 100\,K steps to 778\,K. Subsequent measurements were performed every 6\,K. Data were corrected in the same manner as the room temperature X-ray measurements.

\subsection{Reverse Monte Carlo Modelling}

To produce the structural configuration of the melt quenched glass, The RMCProfile software\cite{tucker_rmcprofile_2007} was used to refine an atomistic starting model for an amorphous, non melt-quenched ZIF described previously,\cite{bennett_structure_2010} against the neutron and X-ray $F(Q)$ and neutron PDF room temperature data. The resultant configuration was then used as a starting input model for RMCProfile, and refined against the variable temperature X-ray data collected, in order to produce the configuration for the liquid ZIF. The density of the model was adjusted to match that determined experimentally for the glass and liquid phases.

\subsection{First principles molecular dynamics}

The behaviour of zeolitic imidazolate frameworks as a function of temperature was studied by means of density functional theory (DFT)-based molecular dynamics (MD) simulations, using the Quickstep module~\cite{van2005} of the CP2K software package.\cite{cp2k} We used the hybrid Gaussian and plane wave method GPW\cite{vandevondele_2005} as implemented in CP2K.  The simulations were performed in the constant-volume $(N, V, T)$ ensemble with fixed size and shape of the unit cell. A timestep of 0.5\,fs was used in the MD runs, the temperature was controlled by velocity rescaling.\cite{bus2007}

The unsually large temperature at which the simulations were performed requires careful fine-tuning of the simulation protocol. In particular, the exchange-correlation energy was evaluated in the PBE approximation,\cite{per1996} and the dispersion interactions were treated at the DFT-D3 level.\cite{gri2010} The Quickstep module uses a multi-grid system to map the basis functions onto. We kept the default number of 4 different grids but chose a relatively high plane-wave cutoff for the electronic density to be 600\,Ry, as already used in Ref.~\citenum{hai2013}, and a relative cutoff (keyword REL$\_$CUTOFF in CP2K) of 40\,Ry for high accuracy. Valence electrons were described by double-zeta valence polarized basis sets and norm-conserving Goedecker--Teter--Hutter\cite{goe1996} pseudopotentials all adapted for PBE (DZVP-GTH-PBE) for H, C and N or optimized for solids (DZVP-MOLOPT-SR-GTH) in the case of Zn.

The unit cell studied for ZIF-4 (space group $Pbca$) is the orthorhombic primitive unit cell, which contains 272~atoms, with cell parameters $a = 15.423$~\AA, $b = 15.404$~\AA, $c = 18.438$~\AA, and $\alpha = \beta = \gamma = 90$\textdegree. Representative input files for the molecular dynamics simulations are available as supporting information, and online in our data repository at {\small\url{https://github.com/fxcoudert/citable-data}}.

\subsection{Trajectory analysis}

The Lindemann ratio $\Delta$ is computed from the width of the first peaks in the different partial radial distribution functions as a measure of the fluctuation of atomic positions and interparticular distances:
\begin{equation}
\Delta = \frac{\text{FWHM}}{d_0}
\end{equation}
where FWHM is the full width at half maximum of the first partial radial distribution function peak (estimated by a Gaussian fit) and $d_0$ corresponds to the mean interatomic distance (calculated as the maximum of the first peak), i.e. $d_0 = 5.95$\,{\AA} for Zn--Zn and $d_0 = 2.0$\,{\AA} for Zn--N.

The coordination number for nitrogen atoms around the zinc cation is computed by taking a cut-off radius of 2.5\,{\AA}, a value chosen from the Zn--N partial radial distribution function at room temperature. We checked that the precise value used does not influence the outcome of the calculations, nor does the choice of a discontinuous criterion (vs. the use of a damping function near the cut-off value).

From the MD simulations, we can extract the internal energy $U(T)$ as the average of the total energy of the system. From these values we can then calculated the heat capacities $C_V(T)$ by finite differences:
\begin{equation}
C_v\left(\frac{T_1+T_2}{2}\right) = \frac{U(T_2)-U(T_1)}{T_2-T_1}
\end{equation}
where $T_1$ and $T_2$ are two successive temperatures in our series of simulations. Furthermore, we can estimate the enthalpy of fusion by writing the energy difference in heating the system from $T_1 = 1,000$\,K to $T_2 = 1,500$\,K:
\begin{equation}
\Delta U = C_{V}(T_1) \times (T_2 - T_1) + \Delta H\e{fus}
\end{equation}

In order to compute the total porous volume, we used the freely available software Zeo++.\cite{pinheiro_characterization_2013, martin_addressing_2012, willems_algorithms_2012} It uses a geometric decomposition of space to compute the accessible and non-accessible volume to a sphere of a given radius. We have taken a value a 1.2\,{\AA} simulating the porous volume as seen by a helium molecule, calculating the distribution of instantaneous total pore space (sum of the accessible and the non-accessible volume) along the MD trajectories at each temperature.

\subsection{Acknowledgements}
We thank Anne Boutin, Alain Fuchs, Anthony Cheetham, and Rodolphe Vuilleumier for fruitful discussions. This work benefitted from the financial support of ANRT (th\`ese CIFRE 2015/0268). We acknowledge access to HPC platforms provided by a GENCI grant (A0010807069). TDB would like to thank the Royal Society for a University Research Fellowship. We also thank Diamond Light Source for access to beamline I15-1 (EE15676), and Dean Keeble and Philip Chater for assistance with data collection on I15-1 during its initial commissioning phase. We gratefully acknowledge the Science and Technology Facilities Council (STFC) for access to neutron beamtime at ISIS on the GEM instrument. This research used resources of the Advanced Photon Source (Beamline 11-ID-B, GUP44665), a U.S. Department of Energy (DOE) Office of Science User Facility operated for the DOE Office of Science by Argonne National Laboratory under Contract No. DE-AC02-06CH11357.

\bibliography{article}

\providecommand{\latin}[1]{#1}
\makeatletter
\providecommand{\doi}
  {\begingroup\let\do\@makeother\dospecials
  \catcode`\{=1 \catcode`\}=2\doi@aux}
\providecommand{\doi@aux}[1]{\endgroup\texttt{#1}}
\makeatother
\providecommand*\mcitethebibliography{\thebibliography}
\csname @ifundefined\endcsname{endmcitethebibliography}
  {\let\endmcitethebibliography\endthebibliography}{}
\begin{mcitethebibliography}{47}
\providecommand*\natexlab[1]{#1}
\providecommand*\mciteSetBstSublistMode[1]{}
\providecommand*\mciteSetBstMaxWidthForm[2]{}
\providecommand*\mciteBstWouldAddEndPuncttrue
  {\def\EndOfBibitem{\unskip.}}
\providecommand*\mciteBstWouldAddEndPunctfalse
  {\let\EndOfBibitem\relax}
\providecommand*\mciteSetBstMidEndSepPunct[3]{}
\providecommand*\mciteSetBstSublistLabelBeginEnd[3]{}
\providecommand*\EndOfBibitem{}
\mciteSetBstSublistMode{f}
\mciteSetBstMaxWidthForm{subitem}{(\alph{mcitesubitemcount})}
\mciteSetBstSublistLabelBeginEnd
  {\mcitemaxwidthsubitemform\space}
  {\relax}
  {\relax}

\bibitem[Mason \latin{et~al.}(2015)Mason, Oktawiec, Taylor, Hudson, Rodriguez,
  Bachman, Gonzalez, Cervellino, Guagliardi, Brown, Llewellyn, Masciocchi, and
  Long]{mason_methane_2015}
Mason,~J.~A.; Oktawiec,~J.; Taylor,~M.~K.; Hudson,~M.~R.; Rodriguez,~J.;
  Bachman,~J.~E.; Gonzalez,~M.~I.; Cervellino,~A.; Guagliardi,~A.;
  Brown,~C.~M.; Llewellyn,~P.~L.; Masciocchi,~N.; Long,~J.~R. \emph{Nature}
  \textbf{2015}, \emph{527}, 357--361\relax
\mciteBstWouldAddEndPuncttrue
\mciteSetBstMidEndSepPunct{\mcitedefaultmidpunct}
{\mcitedefaultendpunct}{\mcitedefaultseppunct}\relax
\EndOfBibitem
\bibitem[Rodenas \latin{et~al.}(2014)Rodenas, Luz, Prieto, Seoane, Miro, Corma,
  Kapteijn, Llabr\'es~i Xamena, and Gascon]{Rodenas2014}
Rodenas,~T.; Luz,~I.; Prieto,~G.; Seoane,~B.; Miro,~H.; Corma,~A.;
  Kapteijn,~F.; Llabr\'es~i Xamena,~F.~X.; Gascon,~J. \emph{Nat Mater}
  \textbf{2014}, \emph{14}, 48--55\relax
\mciteBstWouldAddEndPuncttrue
\mciteSetBstMidEndSepPunct{\mcitedefaultmidpunct}
{\mcitedefaultendpunct}{\mcitedefaultseppunct}\relax
\EndOfBibitem
\bibitem[Yoon \latin{et~al.}(2016)Yoon, Chang, Lee, Hwang, Hong, Lee, Lee,
  Jang, Yoon, Kwac, Jung, Pillai, Faucher, Vimont, Daturi, F\'erey, Serre,
  Maurin, Bae, and Chang]{Yoon2016}
Yoon,~J.~W. \latin{et~al.}  \emph{Nat Mater} \textbf{2016}, 57\relax
\mciteBstWouldAddEndPuncttrue
\mciteSetBstMidEndSepPunct{\mcitedefaultmidpunct}
{\mcitedefaultendpunct}{\mcitedefaultseppunct}\relax
\EndOfBibitem
\bibitem[Mondloch \latin{et~al.}(2015)Mondloch, Katz, Isley~III, Ghosh, Liao,
  Bury, Wagner, Hall, DeCoste, Peterson, Snurr, Cramer, Hupp, and
  Farha]{mondloch_destruction_2015}
Mondloch,~J.~E.; Katz,~M.~J.; Isley~III,~W.~C.; Ghosh,~P.; Liao,~P.; Bury,~W.;
  Wagner,~G.~W.; Hall,~M.~G.; DeCoste,~J.~B.; Peterson,~G.~W.; Snurr,~R.~Q.;
  Cramer,~C.~J.; Hupp,~J.~T.; Farha,~O.~K. \emph{Nature Mater.} \textbf{2015},
  \emph{14}, 512--516\relax
\mciteBstWouldAddEndPuncttrue
\mciteSetBstMidEndSepPunct{\mcitedefaultmidpunct}
{\mcitedefaultendpunct}{\mcitedefaultseppunct}\relax
\EndOfBibitem
\bibitem[Horike \latin{et~al.}(2013)Horike, Umeyama, and
  Kitagawa]{horike_ion_2013}
Horike,~S.; Umeyama,~D.; Kitagawa,~S. \emph{Acc. Chem. Res.} \textbf{2013},
  \emph{46}, 2376--2384\relax
\mciteBstWouldAddEndPuncttrue
\mciteSetBstMidEndSepPunct{\mcitedefaultmidpunct}
{\mcitedefaultendpunct}{\mcitedefaultseppunct}\relax
\EndOfBibitem
\bibitem[Sholl and Lively(2015)Sholl, and Lively]{sholl_defects_2015}
Sholl,~D.~S.; Lively,~R.~P. \emph{J. Phys. Chem. Lett.} \textbf{2015},
  \emph{6}, 3437--3444\relax
\mciteBstWouldAddEndPuncttrue
\mciteSetBstMidEndSepPunct{\mcitedefaultmidpunct}
{\mcitedefaultendpunct}{\mcitedefaultseppunct}\relax
\EndOfBibitem
\bibitem[Cairns and Goodwin(2013)Cairns, and Goodwin]{Cairns2013}
Cairns,~A.~B.; Goodwin,~A.~L. \emph{Chem. Soc. Rev.} \textbf{2013}, \emph{42},
  4881\relax
\mciteBstWouldAddEndPuncttrue
\mciteSetBstMidEndSepPunct{\mcitedefaultmidpunct}
{\mcitedefaultendpunct}{\mcitedefaultseppunct}\relax
\EndOfBibitem
\bibitem[Schneemann \latin{et~al.}(2014)Schneemann, Bon, Schwedler, Senkovska,
  Kaskel, and Fischer]{schneemann_flexible_2014}
Schneemann,~A.; Bon,~V.; Schwedler,~I.; Senkovska,~I.; Kaskel,~S.;
  Fischer,~R.~A. \emph{Chem. Soc. Rev.} \textbf{2014}, \emph{43},
  6062--6096\relax
\mciteBstWouldAddEndPuncttrue
\mciteSetBstMidEndSepPunct{\mcitedefaultmidpunct}
{\mcitedefaultendpunct}{\mcitedefaultseppunct}\relax
\EndOfBibitem
\bibitem[Morris and Čejka(2015)Morris, and Čejka]{Morris2015}
Morris,~R.~E.; Čejka,~J. \emph{Nature Chem} \textbf{2015}, \emph{7},
  381--388\relax
\mciteBstWouldAddEndPuncttrue
\mciteSetBstMidEndSepPunct{\mcitedefaultmidpunct}
{\mcitedefaultendpunct}{\mcitedefaultseppunct}\relax
\EndOfBibitem
\bibitem[Bennett \latin{et~al.}(2017)Bennett, Cheetham, Fuchs, and
  Coudert]{bennett_interplay_2017}
Bennett,~T.~D.; Cheetham,~A.~K.; Fuchs,~A.~H.; Coudert,~F.-X. \emph{Nature
  Chem.} \textbf{2017}, \emph{9}, 11--16\relax
\mciteBstWouldAddEndPuncttrue
\mciteSetBstMidEndSepPunct{\mcitedefaultmidpunct}
{\mcitedefaultendpunct}{\mcitedefaultseppunct}\relax
\EndOfBibitem
\bibitem[Bennett and Cheetham(2014)Bennett, and
  Cheetham]{bennett_amorphous_2014}
Bennett,~T.~D.; Cheetham,~A.~K. \emph{Acc. Chem. Res.} \textbf{2014},
  \emph{47}, 1555--1562\relax
\mciteBstWouldAddEndPuncttrue
\mciteSetBstMidEndSepPunct{\mcitedefaultmidpunct}
{\mcitedefaultendpunct}{\mcitedefaultseppunct}\relax
\EndOfBibitem
\bibitem[Park \latin{et~al.}(2006)Park, Ni, Cote, Choi, Huang, Uribe-Romo,
  Chae, O'Keeffe, and Yaghi]{Park2006}
Park,~K.~S.; Ni,~Z.; Cote,~A.~P.; Choi,~J.~Y.; Huang,~R.; Uribe-Romo,~F.~J.;
  Chae,~H.~K.; O'Keeffe,~M.; Yaghi,~O.~M. \emph{Proceedings of the National
  Academy of Sciences} \textbf{2006}, \emph{103}, 10186--10191\relax
\mciteBstWouldAddEndPuncttrue
\mciteSetBstMidEndSepPunct{\mcitedefaultmidpunct}
{\mcitedefaultendpunct}{\mcitedefaultseppunct}\relax
\EndOfBibitem
\bibitem[Tian \latin{et~al.}(2007)Tian, Zhao, Chen, Zhang, Weng, and
  Zhao]{tian_design_2007}
Tian,~Y.-Q.; Zhao,~Y.-M.; Chen,~Z.-X.; Zhang,~G.-N.; Weng,~L.-H.; Zhao,~D.-Y.
  \emph{Chem. Eur. J.} \textbf{2007}, \emph{13}, 4146--4154\relax
\mciteBstWouldAddEndPuncttrue
\mciteSetBstMidEndSepPunct{\mcitedefaultmidpunct}
{\mcitedefaultendpunct}{\mcitedefaultseppunct}\relax
\EndOfBibitem
\bibitem[Bennett \latin{et~al.}(2015)Bennett, Tan, Yue, Baxter, Ducati,
  Terrill, Yeung, Zhou, Chen, Henke, Cheetham, and
  Greaves]{bennett_hybrid_2015}
Bennett,~T.~D.; Tan,~J.-C.; Yue,~Y.; Baxter,~E.; Ducati,~C.; Terrill,~N.~J.;
  Yeung,~H. H.~M.; Zhou,~Z.; Chen,~W.; Henke,~S.; Cheetham,~A.~K.;
  Greaves,~G.~N. \emph{Nature Comm.} \textbf{2015}, \emph{6}, 8079\relax
\mciteBstWouldAddEndPuncttrue
\mciteSetBstMidEndSepPunct{\mcitedefaultmidpunct}
{\mcitedefaultendpunct}{\mcitedefaultseppunct}\relax
\EndOfBibitem
\bibitem[Umeyama \latin{et~al.}(2015)Umeyama, Horike, Inukai, Itakura, and
  Kitagawa]{umeyama_reversible_2015}
Umeyama,~D.; Horike,~S.; Inukai,~M.; Itakura,~T.; Kitagawa,~S. \emph{J. Am.
  Chem. Soc.} \textbf{2015}, \emph{137}, 864--870\relax
\mciteBstWouldAddEndPuncttrue
\mciteSetBstMidEndSepPunct{\mcitedefaultmidpunct}
{\mcitedefaultendpunct}{\mcitedefaultseppunct}\relax
\EndOfBibitem
\bibitem[MacFarlane \latin{et~al.}(2016)MacFarlane, Forsyth, Howlett, Kar,
  Passerini, Pringle, Ohno, Watanabe, Yan, Zheng, Zhang, and
  Zhang]{MacFarlane2016}
MacFarlane,~D.~R.; Forsyth,~M.; Howlett,~P.~C.; Kar,~M.; Passerini,~S.;
  Pringle,~J.~M.; Ohno,~H.; Watanabe,~M.; Yan,~F.; Zheng,~W.; Zhang,~S.;
  Zhang,~J. \emph{Nat. Rev. Mater.} \textbf{2016}, \emph{1}, 15005\relax
\mciteBstWouldAddEndPuncttrue
\mciteSetBstMidEndSepPunct{\mcitedefaultmidpunct}
{\mcitedefaultendpunct}{\mcitedefaultseppunct}\relax
\EndOfBibitem
\bibitem[Giri \latin{et~al.}(2015)Giri, Del~P{\'o}polo, Melaugh, Greenaway,
  R{\"a}tzke, Koschine, Pison, Gomes, Cooper, and James]{giri_liquids_2015}
Giri,~N.; Del~P{\'o}polo,~M.~G.; Melaugh,~G.; Greenaway,~R.~L.; R{\"a}tzke,~K.;
  Koschine,~T.; Pison,~L.; Gomes,~M. F.~C.; Cooper,~A.~I.; James,~S.~L.
  \emph{Nature} \textbf{2015}, \emph{527}, 216--220\relax
\mciteBstWouldAddEndPuncttrue
\mciteSetBstMidEndSepPunct{\mcitedefaultmidpunct}
{\mcitedefaultendpunct}{\mcitedefaultseppunct}\relax
\EndOfBibitem
\bibitem[Kohara \latin{et~al.}(2014)Kohara, Akola, Patrikeev, Ropo, Ohara,
  Itou, Fujiwara, Yahiro, Okada, Ishikawa, Mizuno, Masuno, Watanabe, and
  Usuki]{Kohara2014}
Kohara,~S.; Akola,~J.; Patrikeev,~L.; Ropo,~M.; Ohara,~K.; Itou,~M.;
  Fujiwara,~A.; Yahiro,~J.; Okada,~J.~T.; Ishikawa,~T.; Mizuno,~A.; Masuno,~A.;
  Watanabe,~Y.; Usuki,~T. \emph{Nat Comms} \textbf{2014}, \emph{5}, 5892\relax
\mciteBstWouldAddEndPuncttrue
\mciteSetBstMidEndSepPunct{\mcitedefaultmidpunct}
{\mcitedefaultendpunct}{\mcitedefaultseppunct}\relax
\EndOfBibitem
\bibitem[Corradini \latin{et~al.}(2016)Corradini, Coudert, and
  Vuilleumier]{Corradini2016}
Corradini,~D.; Coudert,~F.-X.; Vuilleumier,~R. \emph{Nature Chem.}
  \textbf{2016}, \emph{8}, 454--460\relax
\mciteBstWouldAddEndPuncttrue
\mciteSetBstMidEndSepPunct{\mcitedefaultmidpunct}
{\mcitedefaultendpunct}{\mcitedefaultseppunct}\relax
\EndOfBibitem
\bibitem[Wharmby \latin{et~al.}(2015)Wharmby, Henke, Bennett, Bajpe, Schwedler,
  Thompson, Gozzo, Simoncic, Mellot-Draznieks, Tao, Yue, and
  Cheetham]{wharmby_extreme_2015}
Wharmby,~M.~T.; Henke,~S.; Bennett,~T.~D.; Bajpe,~S.~R.; Schwedler,~I.;
  Thompson,~S.~P.; Gozzo,~F.; Simoncic,~P.; Mellot-Draznieks,~C.; Tao,~H.;
  Yue,~Y.; Cheetham,~A.~K. \emph{Angew. Chem. Int. Ed.} \textbf{2015},
  \emph{54}, 6447--6451\relax
\mciteBstWouldAddEndPuncttrue
\mciteSetBstMidEndSepPunct{\mcitedefaultmidpunct}
{\mcitedefaultendpunct}{\mcitedefaultseppunct}\relax
\EndOfBibitem
\bibitem[Bennett \latin{et~al.}(2010)Bennett, Goodwin, Dove, Keen, Tucker,
  Barney, Soper, Bithell, Tan, and Cheetham]{bennett_structure_2010}
Bennett,~T.~D.; Goodwin,~A.~L.; Dove,~M.~T.; Keen,~D.~A.; Tucker,~M.~G.;
  Barney,~E.~R.; Soper,~A.~K.; Bithell,~E.~G.; Tan,~J.-C.; Cheetham,~A.~K.
  \emph{Phys. Rev. Lett.} \textbf{2010}, \emph{104}\relax
\mciteBstWouldAddEndPuncttrue
\mciteSetBstMidEndSepPunct{\mcitedefaultmidpunct}
{\mcitedefaultendpunct}{\mcitedefaultseppunct}\relax
\EndOfBibitem
\bibitem[Mei \latin{et~al.}(2007)Mei, Benmore, and Weber]{mei_structure_2007}
Mei,~Q.; Benmore,~C.~J.; Weber,~J. K.~R. \emph{Phys. Rev. Lett.} \textbf{2007},
  \emph{98}\relax
\mciteBstWouldAddEndPuncttrue
\mciteSetBstMidEndSepPunct{\mcitedefaultmidpunct}
{\mcitedefaultendpunct}{\mcitedefaultseppunct}\relax
\EndOfBibitem
\bibitem[not()]{note_temperatures}
Although the higher temperatures would not be physically relevant for the
  experimental system, they are made necessary by the relatively short times
  explored due to the high computational cost of FPMD, in order to gather
  statistics on relatively rare events and high thermodynamic barriers.\relax
\mciteBstWouldAddEndPunctfalse
\mciteSetBstMidEndSepPunct{\mcitedefaultmidpunct}
{}{\mcitedefaultseppunct}\relax
\EndOfBibitem
\bibitem[Chakravarty \latin{et~al.}(2007)Chakravarty, Debenedetti, and
  Stillinger]{chakravarty_lindemann_2007}
Chakravarty,~C.; Debenedetti,~P.~G.; Stillinger,~F.~H. \emph{J. Chem. Phys.}
  \textbf{2007}, \emph{126}, 204508\relax
\mciteBstWouldAddEndPuncttrue
\mciteSetBstMidEndSepPunct{\mcitedefaultmidpunct}
{\mcitedefaultendpunct}{\mcitedefaultseppunct}\relax
\EndOfBibitem
\bibitem[Kelly(1936)]{kelly1936heats}
Kelly,~K. \emph{US Bur. Mines Bull} \textbf{1936}, \emph{393}\relax
\mciteBstWouldAddEndPuncttrue
\mciteSetBstMidEndSepPunct{\mcitedefaultmidpunct}
{\mcitedefaultendpunct}{\mcitedefaultseppunct}\relax
\EndOfBibitem
\bibitem[Samanta \latin{et~al.}(2014)Samanta, Tuckerman, Yu, and
  E]{Samanta2014}
Samanta,~A.; Tuckerman,~M.~E.; Yu,~T.-Q.; E,~W. \emph{Science} \textbf{2014},
  \emph{346}, 729--732\relax
\mciteBstWouldAddEndPuncttrue
\mciteSetBstMidEndSepPunct{\mcitedefaultmidpunct}
{\mcitedefaultendpunct}{\mcitedefaultseppunct}\relax
\EndOfBibitem
\bibitem[Laage(2006)]{laage_molecular_2006}
Laage,~D. \emph{Science} \textbf{2006}, \emph{311}, 832--835\relax
\mciteBstWouldAddEndPuncttrue
\mciteSetBstMidEndSepPunct{\mcitedefaultmidpunct}
{\mcitedefaultendpunct}{\mcitedefaultseppunct}\relax
\EndOfBibitem
\bibitem[Laage and Hynes(2008)Laage, and Hynes]{laage_molecular_2008}
Laage,~D.; Hynes,~J.~T. \emph{J. Phys. Chem. B} \textbf{2008}, \emph{112},
  14230--14242\relax
\mciteBstWouldAddEndPuncttrue
\mciteSetBstMidEndSepPunct{\mcitedefaultmidpunct}
{\mcitedefaultendpunct}{\mcitedefaultseppunct}\relax
\EndOfBibitem
\bibitem[O'Reilly \latin{et~al.}(2007)O'Reilly, Giri, and James]{OReilly2007}
O'Reilly,~N.; Giri,~N.; James,~S.~L. \emph{Chem. Eur. J.} \textbf{2007},
  \emph{13}, 3020--3025\relax
\mciteBstWouldAddEndPuncttrue
\mciteSetBstMidEndSepPunct{\mcitedefaultmidpunct}
{\mcitedefaultendpunct}{\mcitedefaultseppunct}\relax
\EndOfBibitem
\bibitem[Hasell and Cooper(2016)Hasell, and Cooper]{Hasell2016}
Hasell,~T.; Cooper,~A.~I. \emph{Nat. Rev. Mater.} \textbf{2016}, \emph{1},
  16053\relax
\mciteBstWouldAddEndPuncttrue
\mciteSetBstMidEndSepPunct{\mcitedefaultmidpunct}
{\mcitedefaultendpunct}{\mcitedefaultseppunct}\relax
\EndOfBibitem
\bibitem[Thornton \latin{et~al.}(2016)Thornton, Jelfs, Konstas, Doherty, Hill,
  Cheetham, and Bennett]{thornton_porosity_2016}
Thornton,~A.~W.; Jelfs,~K.~E.; Konstas,~K.; Doherty,~C.~M.; Hill,~A.~J.;
  Cheetham,~A.~K.; Bennett,~T.~D. \emph{Chem. Commun.} \textbf{2016},
  \emph{52}, 3750--3753\relax
\mciteBstWouldAddEndPuncttrue
\mciteSetBstMidEndSepPunct{\mcitedefaultmidpunct}
{\mcitedefaultendpunct}{\mcitedefaultseppunct}\relax
\EndOfBibitem
\bibitem[Keen(2001)]{keen_comparison_2001}
Keen,~D.~A. \emph{J. Appl. Cryst.} \textbf{2001}, \emph{34}, 172--177\relax
\mciteBstWouldAddEndPuncttrue
\mciteSetBstMidEndSepPunct{\mcitedefaultmidpunct}
{\mcitedefaultendpunct}{\mcitedefaultseppunct}\relax
\EndOfBibitem
\bibitem[Sop()]{Soper2011}
Soper, A. K. GudrunN and GudrunX: Programs for Correcting Raw Neutron and X-ray
  Diffraction Data to Differential Scattering Cross Section. Tech. Rep.
  RAL-TR-2011-013 (Rutherford Appleton Laboratory, 2011).\relax
\mciteBstWouldAddEndPunctfalse
\mciteSetBstMidEndSepPunct{\mcitedefaultmidpunct}
{}{\mcitedefaultseppunct}\relax
\EndOfBibitem
\bibitem[Soper and Barney(2011)Soper, and Barney]{soper_extracting_2011}
Soper,~A.~K.; Barney,~E.~R. \emph{J. Appl. Cryst.} \textbf{2011}, \emph{44},
  714--726\relax
\mciteBstWouldAddEndPuncttrue
\mciteSetBstMidEndSepPunct{\mcitedefaultmidpunct}
{\mcitedefaultendpunct}{\mcitedefaultseppunct}\relax
\EndOfBibitem
\bibitem[Tucker \latin{et~al.}(2007)Tucker, Keen, Dove, Goodwin, and
  Hui]{tucker_rmcprofile_2007}
Tucker,~M.~G.; Keen,~D.~A.; Dove,~M.~T.; Goodwin,~A.~L.; Hui,~Q. \emph{J. Phys.
  Condens. Matter} \textbf{2007}, \emph{19}, 335218\relax
\mciteBstWouldAddEndPuncttrue
\mciteSetBstMidEndSepPunct{\mcitedefaultmidpunct}
{\mcitedefaultendpunct}{\mcitedefaultseppunct}\relax
\EndOfBibitem
\bibitem[Vande{V}ondele \latin{et~al.}(2005)Vande{V}ondele, Krack, Mohamed,
  Parrinello, Chassaing, and Hutter]{van2005}
Vande{V}ondele,~J.; Krack,~M.; Mohamed,~F.; Parrinello,~M.; Chassaing,~T.;
  Hutter,~J. \emph{Comput. Phys. Comm.} \textbf{2005}, \emph{167},
  103--128\relax
\mciteBstWouldAddEndPuncttrue
\mciteSetBstMidEndSepPunct{\mcitedefaultmidpunct}
{\mcitedefaultendpunct}{\mcitedefaultseppunct}\relax
\EndOfBibitem
\bibitem[cp2()]{cp2k}
\url{http://www.cp2k.org}\relax
\mciteBstWouldAddEndPuncttrue
\mciteSetBstMidEndSepPunct{\mcitedefaultmidpunct}
{\mcitedefaultendpunct}{\mcitedefaultseppunct}\relax
\EndOfBibitem
\bibitem[VandeVondele \latin{et~al.}(2005)VandeVondele, Krack, Mohamed,
  Parrinello, Chassaing, and Hutter]{vandevondele_2005}
VandeVondele,~J.; Krack,~M.; Mohamed,~F.; Parrinello,~M.; Chassaing,~T.;
  Hutter,~J. \emph{Comput. Phys. Comm.} \textbf{2005}, \emph{167},
  103--128\relax
\mciteBstWouldAddEndPuncttrue
\mciteSetBstMidEndSepPunct{\mcitedefaultmidpunct}
{\mcitedefaultendpunct}{\mcitedefaultseppunct}\relax
\EndOfBibitem
\bibitem[Bussi \latin{et~al.}(2007)Bussi, Donadio, and Parrinello]{bus2007}
Bussi,~G.; Donadio,~D.; Parrinello,~M. \emph{J. Chem. Phys.} \textbf{2007},
  \emph{126}, 014101\relax
\mciteBstWouldAddEndPuncttrue
\mciteSetBstMidEndSepPunct{\mcitedefaultmidpunct}
{\mcitedefaultendpunct}{\mcitedefaultseppunct}\relax
\EndOfBibitem
\bibitem[Perdew \latin{et~al.}(1996)Perdew, Burke, and Ernzerhof]{per1996}
Perdew,~J.~P.; Burke,~K.; Ernzerhof,~M. \emph{Phys. Rev. Lett.} \textbf{1996},
  \emph{77}, 3865--3868\relax
\mciteBstWouldAddEndPuncttrue
\mciteSetBstMidEndSepPunct{\mcitedefaultmidpunct}
{\mcitedefaultendpunct}{\mcitedefaultseppunct}\relax
\EndOfBibitem
\bibitem[Grimme \latin{et~al.}(2010)Grimme, Antony, Ehrlich, and
  Krieg]{gri2010}
Grimme,~S.; Antony,~J.; Ehrlich,~S.; Krieg,~H. \emph{J. Chem. Phys.}
  \textbf{2010}, \emph{132}, 154104\relax
\mciteBstWouldAddEndPuncttrue
\mciteSetBstMidEndSepPunct{\mcitedefaultmidpunct}
{\mcitedefaultendpunct}{\mcitedefaultseppunct}\relax
\EndOfBibitem
\bibitem[Haigis \latin{et~al.}(2013)Haigis, Coudert, Vuilleumier, and
  Boutin]{hai2013}
Haigis,~V.; Coudert,~F.-X.; Vuilleumier,~R.; Boutin,~A. \emph{Phys. Chem. Chem.
  Phys.} \textbf{2013}, \emph{15}, 19049--19056\relax
\mciteBstWouldAddEndPuncttrue
\mciteSetBstMidEndSepPunct{\mcitedefaultmidpunct}
{\mcitedefaultendpunct}{\mcitedefaultseppunct}\relax
\EndOfBibitem
\bibitem[Goedecker \latin{et~al.}(1996)Goedecker, Teter, and Hutter]{goe1996}
Goedecker,~S.; Teter,~M.; Hutter,~J. \emph{Phys. Rev. B} \textbf{1996},
  \emph{54}, 1703--1710\relax
\mciteBstWouldAddEndPuncttrue
\mciteSetBstMidEndSepPunct{\mcitedefaultmidpunct}
{\mcitedefaultendpunct}{\mcitedefaultseppunct}\relax
\EndOfBibitem
\bibitem[Pinheiro \latin{et~al.}()Pinheiro, Martin, Rycroft, Jones, Iglesia,
  and Haranczyk]{pinheiro_characterization_2013}
Pinheiro,~M.; Martin,~R.~L.; Rycroft,~C.~H.; Jones,~A.; Iglesia,~E.;
  Haranczyk,~M. \emph{J. Mol. Graph. Model.} \emph{44}, 208--219\relax
\mciteBstWouldAddEndPuncttrue
\mciteSetBstMidEndSepPunct{\mcitedefaultmidpunct}
{\mcitedefaultendpunct}{\mcitedefaultseppunct}\relax
\EndOfBibitem
\bibitem[Martin \latin{et~al.}(2012)Martin, Smit, and
  Haranczyk]{martin_addressing_2012}
Martin,~R.~L.; Smit,~B.; Haranczyk,~M. \emph{J. Chem. Inf. Model.}
  \textbf{2012}, \emph{52}, 308--318\relax
\mciteBstWouldAddEndPuncttrue
\mciteSetBstMidEndSepPunct{\mcitedefaultmidpunct}
{\mcitedefaultendpunct}{\mcitedefaultseppunct}\relax
\EndOfBibitem
\bibitem[Willems \latin{et~al.}()Willems, Rycroft, Kazi, Meza, and
  Haranczyk]{willems_algorithms_2012}
Willems,~T.~F.; Rycroft,~C.~H.; Kazi,~M.; Meza,~J.~C.; Haranczyk,~M.
  \emph{Micro. Meso. Mater.} \emph{149}, 134--141\relax
\mciteBstWouldAddEndPuncttrue
\mciteSetBstMidEndSepPunct{\mcitedefaultmidpunct}
{\mcitedefaultendpunct}{\mcitedefaultseppunct}\relax
\EndOfBibitem
\end{mcitethebibliography}

\end{document}